\documentstyle[12pt]{article}
\def\O{ {\cal O }}
\def\L{ {\cal L }}

\def\prl#1#2#3{ Phys. Rev. Lett. ${\bf{#1}}$ (#2) #3}

\def\np#1#2#3{ Nucl. Phys. ${\bf{#1}}$ (#2) #3}
\def\zp#1#2#3{ Z. Phys. ${\bf{#1}}$ (#2) #3}

\def\bq{\begin{equation}}
\def\eq{\end{equation}}
\begin{document}
             PM 97-08, april 97 \\
{\ \ }\\[0.3in]
\centerline{\Large
       Constraints on anomalous gauge couplings} 
\centerline{\Large from present LEP1 and future LEP2, BNL data}
{\ \ }\\[0.2in]
\centerline{F.M. Renard$^{\rm a}$, S. Spagnolo$^{\rm b}$ and 
            C. Verzegnassi$^{\rm b}$}
\vskip 0.5cm
\centerline{\small $^{\rm a}$\it Physique Math\'ematique et Th\'eorique, 
            UPRES-A 5032,}
\centerline{\small \it Universit\'e de Montpellier II, 
            F-34095 Montpellier Cedex 5.}
\vskip 0.3cm
\centerline{\small $^{\rm b}$\it Dipartimento di Fisica Universit\`a di 
                   Lecce and INFN, Sezione di Lecce,}
\centerline{\small \it via Arnesano, 73100 Lecce, Italy}

\vskip 1.cm
\begin{abstract}
We analyze, in a rather general model where anomalous 
triple gauge couplings are present, the visible effects
in R$_b$ (measured at LEP1), in W pair
production (to be measured at LEP2) and in the muon anomalous 
magnetic moment (to be measured at BNL). 
From the combination of the three experiments a
remarkable improvement on the pure LEP2 constraints is obtained. 
\end{abstract}
\vspace{2cm}
\newpage

At the end of the high precision measurements performed at LEP1, the 
Standard Model predictions concerning the Z boson-fermion interactions 
have been verified to a level of accuracy that is generally of a few 
permille \cite{PEW}. 
The fact that, within these limits, no signal of any possible kind
of new physics has been detected does not necessarily imply, though, 
that the same conclusion should apply to different sectors of the 
Standard Model. 
This is particularly true for the purely gauge boson interactions whose 
form, in the Standard Model (SM), is severely constrained 
by the assumed requests of 
local gauge invariance and renormalizability. 

The possibility that ``anomalous'' boson 
gauge couplings exist has been, in fact, the subject of a number of 
theoretical speculations \cite{Anomali} and discussions \cite{Eretici} 
that we shall not review in this short letter, particularly since they 
can be already found in excellent recent review papers \cite{LEP2}. 
Here we shall follow the approach that was presented by Hagiwara, 
Ishihara, Szalapski and Zeppenfeld \cite{HISZ}.
It is based on the assumption that such anomalous couplings appear as
residual effects of particles and interactions lying beyond the SM whose
dynamics is not yet known and is generically called new physics (NP).
It is assumed that this dynamics is characterized by a scale $\Lambda$
which is much higher than the electroweak scale, $\Lambda >> M_W$.
It is then reasonable to expect that at energies much lower than
$\Lambda$ these residual effects should preserve the standard
SU(2)$\times$U(1) gauge invariance before spontaneous breaking
occurs through the Higgs mechanism. Consequently these effects 
can be described by the effective lagrangian method \cite{Buch} and
again because of the assumption that $\Lambda >> M_W$
only dimension six operators constructed with standard bosonic fields
will be retained. We further restrict to CP-conserving interactions
and this leaves seven possible operators listed in \cite{HISZ}.
Among them only three contribute to anomalous 3-gauge boson
couplings. They are dubbed 
\begin{eqnarray} 
\O_{WWW} &= & Tr[ \hat{W}^{\ \ \nu}_\mu
  \hat{W}^{\ \ \lambda}_\nu  
  \hat{W}^{\ \ \mu}_\lambda] \ \ \
 ,  \\  
\O_{W} & = & \, (D_\mu \Phi)^\dagger 
\hat W^{\mu \nu} (D_\nu \Phi) \ \ \  , \\
\O_{B} & = & \, (D_\mu \Phi)^\dagger\hat B^{\mu \nu} (D_\nu
\Phi)\ \ \  , \\ 
\end{eqnarray}
\noindent
where
\bq
D_{\mu}  =  \partial_\mu + i~ {g\prime\over2} B_\mu +
i~ \frac{g}{2} \sigma^a   W^a_\mu \\
\eq 
\bq
\hat B^{\mu \nu}= i~ {g\prime\over2} B_{\mu \nu} \ \ \ \ \ \ \ \  
\hat{W}_{ \nu \mu}=i~ \frac{g}{2} \sigma^a   W^a_{ \nu \mu}
\eq
\noindent
so that the effective lagrangian is written
\bq
\L = {1\over\Lambda^2} (f_{WWW}\O_{WWW}+f_{W}\O_{W}+f_{B}\O_{B})
\eq

The general anomalous $\gamma WW$ and $ZWW$ couplings defined in
\cite{Anomali} can be expressed in terms of the three independent
parameters appearing in the above lagrangian. Only five of them are
feeded by the above three operators.
 In the notation of reference \cite{HISZ}, they are:
\begin{eqnarray} 
    {g_1}_Z & = & 1 + f_W \frac{M_W^2}{2\Lambda^2}\\
   \kappa_\gamma & = & 1+(f_B+f_W) \frac{M_W^2}{2\Lambda^2}\\
        \kappa_Z & = & 1+ [f_W-s^2_W(f_B+f_W)] \frac{M_Z^2}{2\Lambda^2}\\
\lambda_\gamma = \lambda_Z  \equiv \lambda & = & 
                               \frac{3M_W^2g^2}{2\Lambda^2} f_{WWW}.
                               \label{lambdaeq}
\end{eqnarray}
 
One of the best places to search for the existence of
anomalous $\gamma WW$ and $ZWW$ couplings
is the process $e^+e^- \to W^+W^-$. This investigation 
is presently starting at LEP2. An analysis of what could be expected
has been previously performed in refs. \cite{LEP2,Kneur}. 
The results of these analyses are
represented by exclusion plots in the space of the three parameters. 
More precisely, one defines there the quantities: 
\begin{eqnarray}
 x & = & f_B\frac{M_W^2}{2\Lambda^2}\\
 y & = & f_W\frac{M_W^2}{2\Lambda^2}
\end{eqnarray} 
(in ref.\cite{LEP2} they are denoted $\alpha_{B\Phi}$ and
$\alpha_{W\Phi}$ respectively)
and fits them together with $\lambda$ defined in eq.(\ref{lambdaeq}). 
In practice, one rather shows three bidimensional figures in the $(x,y)$, 
$(x,\lambda)$ and $(y,\lambda)$ planes representing the regions
inside which $x,y$ and $\lambda$ would be constrained to lie, at 95$\%$ C.L., 
if no visible effects were seen in the final WW channel, that means
that the experimental measurements agree with the SM prediction within
the assumed uncertainty. The solid lines in 
Fig.1-3 show such
3-parameter constraints 
derived for $\sqrt{s}=190$~GeV with an integrated luminosity of 500~pb$^{-1}$
and relying on WW events where at least one W decays leptonically (the
inclusion of events where both W decays hadronically should slightly
improve the constraints, rougly by a factor 1.4, but the precise
analysis has not been done yet \cite{jlk}).

The information on $x,y$ and $\lambda$ obtained by 
analyzing the LEP2 data 
is provided by a reaction where two real Ws are produces by a virtual V.
This is not the only possible source of constraints. 
At least two different
processes can lead to information on the same parameters, 
by an analysis of
special one-loop contributions where the same VWW vertex
appears but \underline{virtual} Ws (and V) are implied. 
We shall now list
these two cases and consider them in some details.

\vskip 0.3cm
I) \underline{\large Measurements of the ratio 
R$_b=\Gamma_{{\rm Z}\rightarrow {\rm b}\bar{\rm b}}/
     \Gamma_{{\rm Z}\rightarrow {\rm hadrons}}$ on top of} \newline
   \underline{\large the Z$^0$ resonance}. \par
It has been recently emphasized \cite{RV,Spagnoli} that the value of the
ratio R$_b$ would be substantially affected by those anomalous couplings
that modify the Zb$\bar{\rm b}$ vertex by virtual effects 
proportional to m$_t^2$. Such an m$_t^2$ enhancement is present when
a "longitudinal" $W_L$ component (also called "would-be-goldstone"
component $H$) is involved in a loop together with a top quark 
and the $Htb$ vertex. 
This already occurs in the SM and it can also occur 
through the operators
$\O_{W}$ and $\O_{B}$ which generate "anomalous"
ZWH and ZHH vertices. The conclusion of ref.\cite{RV}, 
to which we defer for a complete discussion, is that the shift
produced in R$_b$ by the considered model of anomalous gauge couplings
(AGC) would be:
\begin{equation}
\Delta {\rm R}_b^{\rm AGC} = -\frac{4}{6} \:\frac{1+b}{1+b^2} 
                          \left( \frac{g^2}{128\pi^2}\right)
                          \left( \frac{{\rm m}_t^2}{{\rm M}_W^2} \right)
                          \left( 7y -\frac{s^2_W}{c^2_W} x \right)
                         \ln \frac{\Lambda^2}{{\rm M}_z^2}
\end{equation}
where $b=1-\frac{4}{3}s^2_W$. For m$_t^2=175$~GeV and 
$\Lambda=1$~TeV this
gives  numerically:
\begin{equation}
\Delta {\rm R}_b^{\rm AGC} \simeq-0.04 ( y-0.04 x) \label{Rbeq} 
\end{equation}
showing that this effect is, in practice, providing 
constraints on $y$ alone. 

The numerical exploitation of eq.(\ref{Rbeq}) is 
obviously very strongly
dependent on the experimental value of R$_b$, as we fully attribute 
$\Delta {\rm R}_b\equiv $R$_b^{\rm exp}-$R$_b^{\rm SM}$ to 
$\Delta {\rm R}_b^{\rm AGC}$. Both the central value and the 
uncertainty of the experimental measurement are important for 
the conclusion that we want to draw. As there have been recently
some changes in the experimental results we will be careful
about these points. 
If one uses the last (to our knowledge) 
officially communicated averaged
LEP1/SLC value, that reads  \cite{PEW},\cite{Varsaw}
\begin{equation}
{\rm R}_b = 0.2178 \pm 0.0011
\end{equation}
and the SM prediction for $m_t=177~GeV$
\begin{equation}
{\rm R}_b(SM) = 0.2158
\end{equation}
one obtains
\bq 
\Delta {\rm R}_b^{\rm AGC}=0.0020\pm 0.0011
\eq
and  from eq.(\ref{Rbeq}), assuming  $\Lambda=1~TeV$,
we derive a first bound on $y$.
Combined with the LEP2 constraint at 95$\%$ C.L. the resulting domains
are shown in Fig.1a-3a (dotted lines).\par
In order to show the consequences of recent changes in $R_b$
results on our domains we also make an illustration only using
the most recent value published by ALEPH \cite{ALEPH}, which fully
agrees with the SM prediction
\begin{equation}
{\rm R}_b = 0.2159 \pm 0.0009\pm0.0011
\end{equation}
We obtain
\bq 
\Delta {\rm R}_b^{\rm AGC}=0.0001\pm 0.0020
\eq
and the domains shown in Fig.1b-3b (dotted lines). 
As one could expect, the results essentially differ by a shift of
the domain concerning the parameter $y$. However,
because of the importance of the experimental uncertainty the
order of magnitude of the constraints is rather similar  
to the one obtained in Fig.1a-3a.\par
One should also note that our results
depend through eq.(\ref{Rbeq}) on the assumed value
of the new physics scale $\Lambda$, taken as $1~TeV$ in the
illustrations. This is a typical feature of 1-loop effects. The
dependence in $\Lambda$ is only logarithmic and for example increasing
(which would seem more resonable than decreasing)
$\Lambda$ by a factor of two would only affect the result by 
making the constraint roughly a relative
twenty percent more stringent.\par 
In conclusion of this first part one can say that the net effect of 
adding the LEP1 constraint from R$_b$ to
the LEP2 bounds  corresponds to a strong
reduction of the limits on $y$, while the two other 
parameters are essentially unaffected. 
 
\vskip 0.3cm
II) \underline{\large (Future) measurement of the muon's anomalous magnetic}
    \newline
    \underline{\large moment at BNL}. \par
Naively, one would expect that anomalous $\gamma$WW couplings could affect 
to a sensible extent the measured value of the muon's anomalous magnetic moment
$a_\mu=\frac{1}{2}(g-2)$. In fact in the Standard Model, 
the contribution of
the graph involving the $\gamma WW$ vertex and neutrino exchange
is rather large as 
compared to the expected future precision of the planned
measurement at BNL \cite{BNL}. 

To be more precise, a few details must be added
at this point. The last experimental value of $a_\mu$ is \cite{FarleyPicasso}:
\begin{equation}
a_\mu^{\rm exp} = 1\:165\:923(8.5)\times 10^{-9}
\end{equation}
not in disagreement with the last available theoretical estimate \cite{tau}:
\begin{equation}
a_\mu^{\rm th}  = 1\:165\:917(1.0)\times 10^{-9}.
\end{equation}
Most of the theoretical uncertainty (88$\%$ of the quadratic error) arises from
the lowest order hadronic contribution to $a_\mu$, which is inferred, through 
a dispersion relation, from data of $e^+e^-$ annihilation into hadronic final
states and of hadronic $\tau$ decays. 
It is expected that the forthcoming BNL experiment will reduce the experimental
error by more than one order of magnitude, i.e.:
\begin{equation}
\delta a_\mu^{\rm exp} = 0.4 \times 10^{-9}.
\end{equation}
It should also be mentioned that an extra reduction of the theoretical error,
coming from new measurements of the total cross section of $e^+e^-$
annihilation in the very low energy region, would be, in principle, in the
reach of the DA$\Phi$NE activities \cite{Franzini}. 
In fact, the $\rho$ resonance region is responsible for more than 50$\%$ of the
quadratic error on the lowest order hadronic contribution to $a_\mu$
\cite{tau}. 
A total integrated luminosity of less than 500~nb$^{-1}$, which requires
fourteen hours of running at a luminosity of 10$^{30}$~cm$^{-2}$s$^{-1}$ 
(more than two orders of magnitude lower than the expected peak luminosity 
of DA$\Phi$NE), should allows to measure $\sigma(e^+e^-\rightarrow \pi^+\pi^-$) 
from threshold for hadro-production up to 1.08~GeV with enough 
precision to push down the error on the hadronic contribution to $a_\mu$ 
from this energy region to the value of $1.5\times 10^{-10}$. 
This would lead, according to \cite{tau}, to a 30$\%$ reduction of the error 
on the total hadronic contribution at one loop and, hence, to a 20$\%$ 
reduction of the current overall theoretical uncertainty. 
Thus, one can imagine that 
$a_\mu$ will be soon measured to a relative accuracy of less than a part per 
million. In this picture, the numerical value of the $\gamma WW$ vertex
contribution is,
in the Standard Model, more than 4 times 
larger that the expected precision on $a_\mu$. 
This means that not unfairly small anomalous 
contributions seem to be in a good shape 
to produce visible effects. 

In fact, this feeling is verified in a number of
rigorous calculations \cite{previous} that were performed some time ago, 
mostly in the unitary gauge. The various results produce a leading term, on
which the agreement is general, and finite terms that depend essentially on the
used regularization scheme and do not always coincide. For this reason, but
also in order to present an alternative calculation of the gauge invariant
quantity $a_\mu$, we have redone the calculation in the Feynmann 't~Hooft 
$\xi=1$ gauge within the dimensional regularization framework. 
The various technical details of the calculation are fully
illustrated elsewhere \cite{SS}. Using the generally accepted prescription 
that sets the formal correspondence 
$\frac{2}{n-4}\rightarrow \ln \frac{\Lambda^2}{\mu^2}$, we arrived to 
the following result:
\begin{equation}
\Delta a_\mu^{\rm AGC} = \frac{G_F}{8\sqrt{2}\pi^2} \: m_\mu^2 
                          2 \: \left[\:\Delta k_\gamma \:\left( \frac{1}{2}   
       - \ln \frac{\Lambda^2}{M_W^2} \right)\: - \lambda_\gamma 
                                   \:\right]
\label{g-2eq}
\end{equation}
where $\Delta k_\gamma = k_\gamma-1$ is equal, in our notation, to $x+y$.  

A few comments on eq. (\ref{g-2eq}) are, at this point, appropriate. First of
all, the leading logarithmic term is in agreement, as expected, with all 
previous calculations. Concerning the extra finite terms, that multiplying
$\lambda$ turns out to be in agreement with the corresponding calculation of
the second of ref. \cite{previous} that was  performed using dimensional
regularization; this does not agree with the analogous terms computed using
different regularization techniques, as exhaustively discussed in the second of
ref. \cite{previous}. The finite term that 
multiplies $\Delta k_\gamma$, which
is non zero in our calculation, would be zero in the first of ref. 
\cite{previous} (unitary gauge and cutoff). Again, this is not particularly
surprising given the quite different regularization prescriptions. 
Since for $\Lambda=1$~TeV, which is generally considered as the
lowest acceptable value for the new physics scale, the relative size of
this term is one order of magnitude smaller than that of the leading 
logarithmically divergent contribution, we have decided to ignore it as a first
quite reasonable approximation, keeping in mind the fact that, 
for (unlikely) smaller
$\Lambda$ values, this might be a slightly arbitrary attitude. 
For what concerns the finite contribution coming from $\lambda$, we 
decided to retain it in a first time and to try to quantify its possible
(regularization scheme dependent) role. Indeed, we have verified, as expected,
that the effect of the (ambiguous) finite contribution from $\lambda$ is
completely negligible. Therefore, at least in this special example, one can
safely ignore it in eq.(\ref{g-2eq}). 
With these prescriptions, we have finally taken 
eq.(\ref{g-2eq}) without the
finite terms and assumed no deviations from the Standard Model prediction
and an overall (conservative) accuracy of $1\cdot 10^{-9}$.
This constraint is then added to the previous 
LEP1, LEP2 constraints. The results of our analysis are shown 
in Fig.1-3(a,b) (starred lines) at 95$\%$ C.L.. 
From inspection of those figures the 
following feature of our analysis should
be underlined: 

a) for what concerns $y$, the addition of the 
$a_\mu$ input has a marginal effect of less than $10\%$;

b) The region allowed for $\lambda$ which is practically completely
derivable from LEP2 is only reduced
by a relative factor of about $20\%$; 

c) in the case of $x$, a much more drastic reduction is 
induced by $a_\mu$, numerically equal to a relative factor of about 3; 

d) the inclusion of $a_\mu$ makes the overall picture 
nicely uniform in the
space of the three parameters $x,y$ and $\lambda$. In fact, 
from the combined analysis, the 
allowed intervals for $x,y$ and $\lambda$ are rather similar and, typically, of
size not larger than about 0.2-0.3 (in modulus). This should be
compared with the present bounds obtained from Tevatron analyses 
\cite{teva} that provide allowed ranges typically one order of
magnitude larger.

In conclusion, we have seen that the information 
on possible anomalous gauge
couplings that is already provided by the existing 
measurement of R$_b$ at
LEP1, and that will be soon improved by the current 
measurements of WW
production at LEP2, would be remarkably enriched by 
the addition of the foreseen
improved determination of the muon anomalous magnetic 
moment at BNL.
Should a
deviation from the Standard Model values be present
with a size, typically, of an ``acceptable'' twenty percent in 
the relevant couplings, it should not
escape an accurate simultaneous overall fit. 
More generally, one should probably insist on the fact that
the direct (tree level) tests provided by LEP2 
do not have the same theoretical base as the
indirect (1-loop) tests provided by LEP1 and by BNL. 
Any discrepancy between
these two types of measurements would constitute a signal for another
kind of virtual effects and stimulate further developments of
high precision tests.

\section*{Acknowledgements}
We would like to thank J.L. Kneur, who made available the 
fit program used for producing the exclusion plots that we present.

\newpage

\section*{Figure captions}
\begin{itemize}

\item[Fig. 1]
Constraints on the $x$ and $y$ anomalous couplings resulting from 
LEP2 forthcoming data (solid line), from the combined LEP2 and  
LEP1 R$_b$ data (dotted line) and from the overall 
information provided by 
LEP2, R$_b$ and future $a_\mu$ measurements at BNL (starred line). 
The LEP2 sensitivity limit refers to a center of 
mass energy $\sqrt{s}=190$~GeV 
and to an integrated luminosity of 500~pb$^{-1}$. 
The supposed precision on 
the muon anomalous magnetic moment is $\delta a_\mu=1\cdot 10^{-9}$. 
The contours shown here represent the projection of 
a threedimensional region 
in the $x,y$ and $\lambda$ space allowed, at 95$\%$ 
C.L., by the experiments. The LEP1 $R_b$ data are taken\\
 (a)  from the Varsaw compilation, ref.\cite{PEW},\cite{Varsaw}\\
 (b)  from ALEPH, ref.\cite{ALEPH}.

\item[Fig. 2]
Projection in the $x-\lambda$ plane of the region allowed, at 95$\%$ C.L., to
the three anomalous couplings present in the considered model by the
combination of present LEP1 and future LEP2, BNL data.  
The meaning of the different curves is the same as in Fig.1a,b.

\item[Fig. 3]
Projection in the $y-\lambda$ plane of the allowed region, 
at 95$\%$ C.L., for the $x,y$ and $\lambda$ parameters from present LEP1 
and future LEP2, BNL data.  
The meaning of the different curves is described in Fig.1a,b.

\end{itemize}


\newpage

\begin{thebibliography}{5}
\bibitem{PEW} A. B\"ohm, ``Results from the Measurements of Electroweak
              Processes at LEP1'', invited talk at the XXXIInd Rencontre 
              de Moriond, Les Arcs/Savoie France, March 15-22, 1997
\bibitem{Varsaw} A. Blondel, ICHEP96 (plenary talk), Warsaw, July
1996
\bibitem{Anomali} K. Gaemers and G. Gounaris, Z. Phys. C1 (1979) 259; 
        \newline  K. Hagiwara, K. Hikasa, R. Peccei and D. Zeppenfeld, 
                  Nucl. Phys. {\bf B282} (1987) 253; 
        \newline  P. Mery, M. Perrottet and F.M. Renard, Z. Phys. 
                  {\bf C36} (1988) 579 
\bibitem{Eretici} A. De R\'ujula, M.G. Gavela, P. Hernandez and 
                  E. Mass\'o, Nucl. Phys. {\bf B384} (1992) 3
\bibitem{LEP2} G. Gounaris, J.L. Kneur, D. Zeppenfeld {\it et al.} 
               ``Triple Gauge Boson couplings'', in ``Physics at LEP2'', 
               G. Altarelli and F. Zwirner eds., CERN Report 1996
\bibitem{HISZ} K. Hagiwara, S. Ishihara, R. Szalapski, D. Zeppenfeld, 
               Phys. Rev. {\bf D48} (1993) 2182
\bibitem{Buch} W. Buchm\"{u}ller and D. Wyler,
\np{B268}{1986}{621}; C.J.C. Burgess and H.J. Schnitzer,
\np{B228}{1983}{454}; C.N. Leung, S.T. Love and S. Rao
\zp{C31}{1986}{433}
\bibitem{Kneur} M. Bilenky, J.L. Kneur, F.M. Renard, D. Schildknecht, 
                Nucl. Phys. {\bf B409} (1993) 22 
\bibitem{jlk} J.L. Kneur, private communication
\bibitem{RV} F.M. Renard and C. Verzegnassi, Phys. Lett. {\bf B345} (1995) 500
\bibitem{Spagnoli} O.J.P. Eboli, S.M. Lietti, M.C. Gonzalez-Garcia, S.F.Novaes, 
                   Phys. Lett. {\bf B339} (1994) 119
\bibitem{ALEPH} The ALEPH collaboration, CERN-PPE/97-018 (1997).
\bibitem{BNL}  V.W. Hughes {\it et al.}, ``The anomalous magnetic moment of 
               the muon'' Bonn 1990, Proceedings, High 
               energy spin physics, vol. 1, 367-382;
    \newline   B.L. Roberts, ``The new muon (g-2) experiment at 
	         Brookhaven'', Heidelberg 1991, Proceedings, The future 
	         of muon physics 101-108, Z. Phys. {\bf C56} (1992) Suppl. 101-108
\bibitem{FarleyPicasso} F.J.M. Farley, E. Picasso, `` The muon g-2 
                        experiments'', In T. Kinoshita, (ed.): 
                        Quantum electrodynamics 479-559
\bibitem{tau} R. Alemany, M. Davier and A. H\"ocker, LAL 97-02 (1997);
    \newline  see also S. Eidelman, F. Jegerlehner, Z. Phys. {\bf C67} 
                       (1995) 585
\bibitem{Franzini}P. Franzini, ``The muon gyromagnetic ratio 
		      and R$_h$ at DA$\Phi$NE'', The Second DA$\Phi$NE Physics 
		      Handbook, INFN-LNF (1995) vol. 2, 471
\bibitem{previous} P. M\'ery, S.E. Moubarik, M. Perrottet, F.M. Renard, 
                   Z. Phys. {\bf C46} (1990) 229;
         \newline  F. Boudjema, K. Hagiwara, C. Hamzaoui, K. Numata, 
                   Phys. Rev. {\bf D43} (1991) 2223; 
         \newline  C. Arzt, M.B. Einhorn and J. Wudka, Phys. Rev. 
                   {\bf D49} (1994) 1370 
         \newline and references therein 
\bibitem{SS} S. Spagnolo, ``Non Standard Electroweak Contributions to 
             the muon g-2 from Anomalous Gauge Boson Couplings'', 
             INFN/TH-97/01 (1997) 

\bibitem{teva} F. Abe et al, \prl{75}{1995}{1017}; 
\newline   S.Abachi et al \prl{75}{1995}{1023}, \prl{77}{1996}{3303} 


\end{thebibliography}
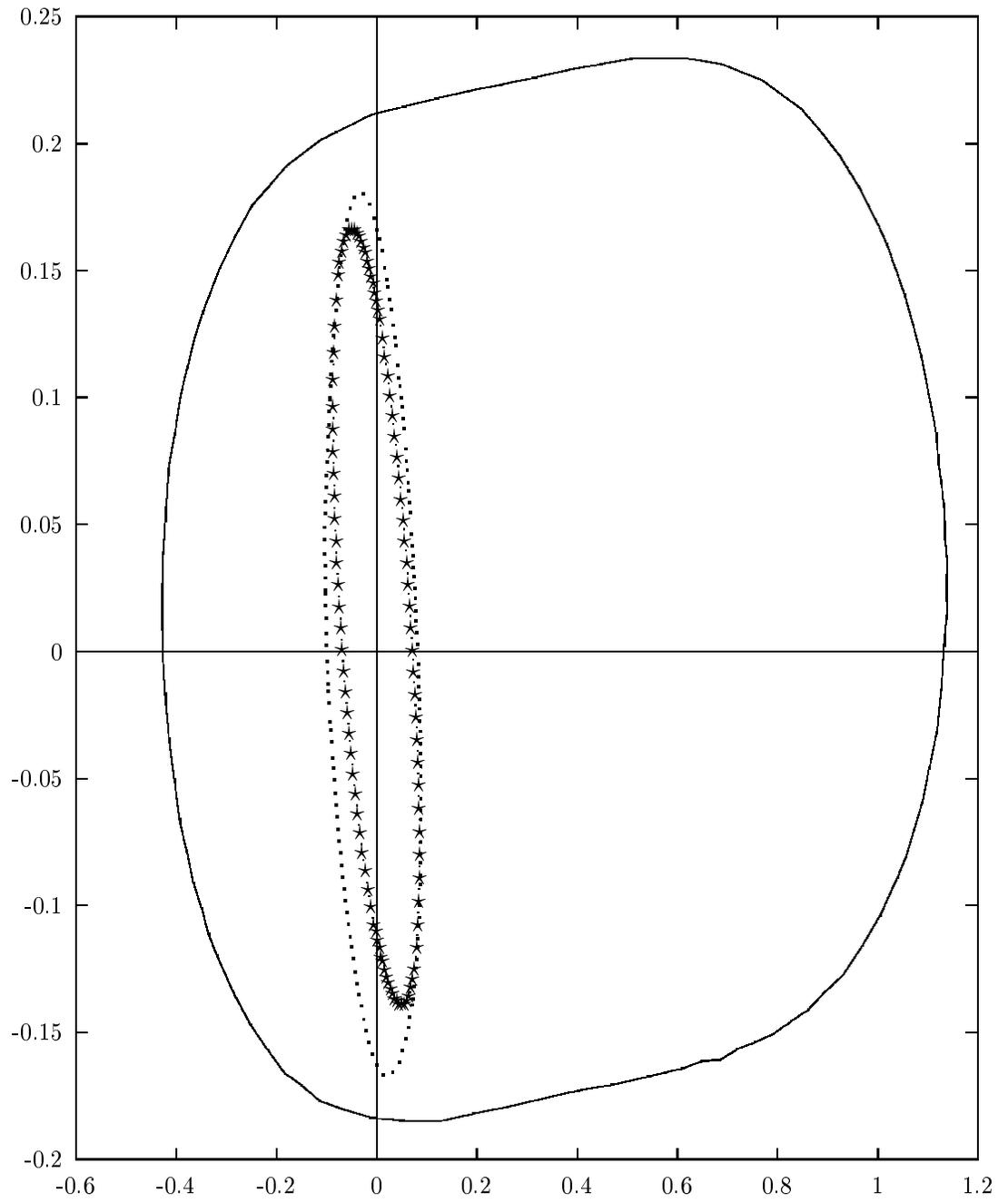
\end{document}